\documentclass[twocolumn,prl,aps,superscriptaddress,showpacs,amsmath,amssymb,floatfix,shortbibliography]{revtex4-1}
\usepackage{graphicx}
\usepackage{amssymb}
\usepackage{amsmath}
\usepackage{epsfig}
\usepackage{color}
\usepackage{mathtools}
\usepackage[colorlinks,linkcolor=blue,anchorcolor=blue,citecolor=blue,urlcolor=blue]{hyperref}

\begin{document}

\title{Exact scaling of geometric phase and fidelity susceptibility and their breakdown across the critical points}
\author{Jia-Ming Cheng}
\affiliation{Key Lab of Quantum Information, CAS, University of Science and Technology of China, Hefei, 230026, P.R. China}
\author{Ming Gong}
\email{gongm@ustc.edu.cn}
\affiliation{Key Lab of Quantum Information, CAS, University of Science and Technology of China, Hefei, 230026, P.R. China}
\affiliation{Synergetic Innovation Center of Quantum Information and Quantum Physics, University of Science and Technology of China, Hefei, 230026, P.R. China}
\author{Guang-Can Guo}
\affiliation{Key Lab of Quantum Information, CAS, University of Science and Technology of China, Hefei, 230026, P.R. China}
\affiliation{Synergetic Innovation Center of Quantum Information and Quantum Physics, University of Science and Technology of China, Hefei, 230026, P.R. China}
\author{Zheng-Wei Zhou}
\email{zwzhou@ustc.edu.cn}
\affiliation{Key Lab of Quantum Information, CAS, University of Science and Technology of China, Hefei, 230026, P.R. China}
\affiliation{Synergetic Innovation Center of Quantum Information and Quantum Physics, University of Science and Technology of China, Hefei, 230026, P.R. China}
\date{\today }

\begin{abstract}
  It was shown via numerical simulations that geometric phase (GP) and fidelity susceptibility (FS) in some quantum models exhibit universal scaling laws 
  across phase transition points. Here we propose a singular function expansion method to determine their exact form across the critical points as well as their 
  corresponding constants. For the models such as anisotropic XY model where the energy gap is closed and reopened at the special points ($k_0 = 0, \pi$), scaling 
  laws can be found as a function of system length $N$ and parameter deviation $\lambda - \lambda_c$ (where $\lambda_c$ is the critical parameter). 
  Intimate relations for the coefficients in GP and FS have also been determined. However in the extended models where the  gap is not closed and reopened at these 
  special points, the scaling as a function of system length $N$ breaks down. We also show that the second order derivative of GP also exhibits some intriguing scaling
  laws across the critical points. These exact results can greatly enrich our understanding of GP and FS in the characterization of quantum phase transitions. 
\end{abstract}


\maketitle

Ever since its theoretical discovery\cite{berry1984quantal}, geometric phase (GP) has permeated into different branches of physics, including
ultracold atoms\cite{ruseckas2005non, atala2013direct, aidelsburger2015measuring}, quantum computation\cite{zhu2003unconventional, duan2001geometric, 
jones2000geometric, ekert2000geometric}, condensed matter physics\cite{avron1983homotopy, resta1994macroscopic, simon1983holonomy, thouless1982quantized, arovas1984fractional} and even 
chemistry physics\cite{juanes2005theoretical, wu1993prediction, mead1992geometric}. GP has become a central concept in amounts of investigations in recent 
decades as an important tool to study the geometric feature of Hamiltonians\cite{brody2001geometric, ashtekar1999geometrical, santamato1985gauge}; especially, it can even be used 
to characterize topological phase transitions\cite{qi2011topological, hasan2010colloquium,xiao2010berry}, which are beyond the accessibility of Landau theory of phase transition. 
This phase can also be directly measured in experiments\cite{bhandari1988observation, ericsson2005measurement, leek2007observation, atala2013direct, xiao2015geometric}. 
Across the critical points the derivative of GP exhibits universal scaling laws\cite{carollo2005geometric, zhu2006scaling, zhu2008geometric}. 

Fidelity susceptibility (FS) based on the overlap of ground state functions is another way beyond the Landau paradigm to characterize quantum phase transitions\cite{quan2006decay,
zanardi2007mixed, gu2010fidelity, chen2007fidelity,you2007fidelity,cozzini2007quantum,buonsante2007ground,schwandt2009quantum,abasto2008fidelity, shan2014scaling, shan2014scaling2, 
chen2008intrinsic, jafari2016quench, wang2013fidelity}. The FS is not defined along a closed trajectory in parameter space, thus it is not directly related to the global geometric 
feature of the ground state. However since the structure of wave functions in two different phases are different, we see that FS also exhibits some scaling laws across critical 
points;  see reviews in Refs. \cite{gu2010fidelity, zhu2008geometric}. 

In previous literatures all these scaling laws are exploited by numerical simulations, thus our understanding of these laws are limited although they have been widely 
investigated\cite{carollo2005geometric, zhu2006scaling, zhu2008geometric, quan2006decay, zanardi2007mixed, gu2010fidelity, chen2007fidelity, you2007fidelity, cozzini2007quantum, 
buonsante2007ground, schwandt2009quantum, abasto2008fidelity, shan2014scaling, shan2014scaling2, chen2008intrinsic, jafari2016quench,wang2013fidelity}. Here in this work these scaling laws are 
obtained exactly using a singular function expansion method, in which all coefficients are also determined exactly. We show that these two measurements are essentially determined by the same physics across the 
critical points, thus their coefficients also have some intimate relations. The coefficients of the divergent terms only reflect how and in which way the energy gap is closed and reopened 
during phase transition, thus do not carry information about the topological properties of ground state wave functions. We also find that the constant term in 
FS is accompanied by a discontinuous jump across the critical points, thus does not have a universal scaling form. 
Finally we show that when the gap is closed and reopened not at these special points ($k_0 = 0, \pm \pi$), 
no scaling laws can be found in both quantities as a function of system length, and the intimate relations between the coefficients in GP and FS also break down. 
These exact results can provide new insights into the characterization of quantum phase transitions using GP and FS.

{\it Basic Method}. We illustrate the basic idea using the following anisotropic XY model\cite{maziero2010quantum,zanardi2006ground,bunder1999effect,lieb1961two, pfeuty1970one},
\begin{equation}
          H=-\sum^{M}_{j=-M} (\frac{1+\gamma}{2}\sigma_{j}^{x}\sigma_{j+1}^{x}+\frac{1-\gamma}{2}\sigma_{j}^{y}\sigma_{j+1}^{y}+\lambda\sigma_{j}^{z}),
        \label{eq-XY}
\end{equation}
where $\lambda$ is the Zeeman field, $\gamma$ is the anisotropy in $x$-$y$ plane and $N = 2M+1$ is the total number of sites. This model reduces to the transverse Ising model
when $\gamma = \pm 1$. To define the geometry phase a circuit of the Hamiltonian is constructed as following, $H_{\phi}=\mathcal{R}_\phi^\dagger H \mathcal{R}_{\phi}$, 
where $\mathcal{R}_{\phi}=\prod_{j=-M}^{M}\exp(i\phi\sigma_{j}^{z}/2)$\cite{carollo2005geometric, zhu2008geometric, zhu2006scaling}. 
Hamiltonian $H_{\phi}$ can be diagonalized by the standard Jordan-Wigner transformation and Bogoliubov transformation\cite{lieb1961two, pfeuty1970one, sachdev2007quantum}, 
and the Bogoliubov-de Gennes (BdG) equation reads as
\begin{equation}
    H_\text{BdG}=\sum_k \Phi_k^\dagger 
    \begin{pmatrix}
        \epsilon_{k}  & \Delta_k  \\
        \Delta_k^*    & - \epsilon_{-k}
    \end{pmatrix} \Phi_k,
    \label{eq-bdg}
\end{equation}
where $\epsilon_k = \lambda - \cos(k)$, $\Delta_k = -ie^{-2i\phi}\gamma \sin(k)$ and $\Phi_k^\dagger = (c_k^\dagger, c_{-k})$ in the Nambu basis with $c_k^\dagger$ 
being the fermion creation operator. The corresponding ground state wave function is written as
\begin{eqnarray}
   \vert g\rangle=\prod_{k>0}(\cos(\frac{\theta_{k}}{2})  + ie^{-i2\phi}\sin(\frac{\theta_{k}}{2}) c_k^\dagger c_{-k}^\dagger)|0\rangle,
    \label{eq-g}
 \end{eqnarray}
where the relative phase is defined by,
\begin{equation}
    \cos\theta_k = {\epsilon_k \over \xi_k}, \quad \sin\theta_k = {i e^{2i\phi} \Delta_k \over \xi_k}, 
    \label{eq-theta}
\end{equation}
and (half of) the energy gap $\xi_k  = \sqrt{\epsilon_k^2 + |\Delta_k|^2}$. With this ground state the GP is determined
\cite{carollo2005geometric, zhu2008geometric, zhu2006scaling},
\begin{eqnarray}
   \Psi_{g}=-\sum_{k>0}\frac{\pi}{M}(1-\cos\theta_{k}),
\end{eqnarray}
which can be regarded as summation of all solid-angles for a ${1 \over 2}$-spin in ''magnetic field'' ${\bf B} = (\Re\Delta_k, \Im\Delta_k, \epsilon_k)$\cite{berry1984quantal}. 
This phase acquired by a closed loop in the parameter space has topological origin\cite{carollo2005geometric} and is 
robust against noise\cite{niskanen2007quantum, falci2000detection}. We are mainly interested in the derivative of the GP across critical points,
which can be defined as,
\begin{eqnarray}
  \label{eq-Psi_g}
  \frac{d\Psi_{g}}{d\lambda}=\frac{\pi}{M}\sum_{k>0}\frac{1}{\xi_k}(1-\frac{\epsilon_{k}^2}{\xi_{k}^2}).
\end{eqnarray}

We first consider the scaling of Eq. \ref{eq-Psi_g} at the critical point as a function of system length $N$. In this model the gap is closed at $k_0 = 0$ $(\pi)$ when $\lambda_c= 1$ $(-1)$ 
and is independent of the anisotropy $\gamma$. These two points are hereafter defined as special points to discriminate them from the case in the extended models discussed 
below. Near the critical point when $\lambda_c = +1$, $\lim_{k\rightarrow 0} \xi_k = |\gamma k|$, thus we have 
the following singular function expansion, which is the key mathematical technique used in this work, 
\begin{equation}
        {1 \over \xi_k} = \chi_k + \mathcal{L}_{\lambda}(k), \quad \chi_k = {1 \over |\gamma|k}.
        \label{eq-Singular}
\end{equation}
Here $\mathcal{L}_{\lambda}(0) = 0$ and $\mathcal{L}_{\lambda}(k)$ is finite everywhere in the whole parameter regime. This divergence fully reflects the linear 
closing and reopening of energy gap across the critical point. The first term is the well-known harmonic number and in the large $N$ limit, 
$\frac{\pi}{M}\sum_{k>0} {1 \over |\gamma|k} \rightarrow {1 \over |\gamma|} (1+{1\over N}) (\Gamma - \ln2 + \ln N)$, where $\Gamma=0.5772...$ is the Euler-Mascheroni 
constant. The remained part converges very fast and in the large $N$ limit can be expressed as an integration,
\begin{eqnarray}
  \mathcal{C} 
  = \int_0^\pi dk [\frac{1}{\xi_k}(1-\frac{\epsilon_{k}^2}{\xi_{k}^2}) -\chi_k]
  =\frac{\ln4|\gamma|}{|\gamma|}-\frac{1+\ln\pi}{|\gamma|}.
\end{eqnarray}
Collecting all these results yields $\frac{d\Psi_{g}}{d\lambda}|_{\lambda = \lambda_c} = \alpha_1 \ln N + \beta_1 + \cdots$, where
\begin{equation}
        \alpha_1 = {1 \over |\gamma|}, \quad \beta_1 = {\Gamma -\ln2 \over |\gamma|} + \frac{\ln4|\gamma|}{|\gamma|}-\frac{1+\ln\pi}{|\gamma|}.
\end{equation}
From the harmonic number we see that the next leading term is ${1 \over |\gamma|} {\ln N \over N}$.

In the thermodynamic limit where the summation of $k$ can be replaced by an integration over the whole momentum space, we try to 
study the scaling law of GP as a function of deviation $\delta \lambda = \lambda - 1$ (for $\lambda_c = + 1$). We need a slightly different
singular function,
\begin{eqnarray}
  \frac{d\Psi_{g}}{d\lambda}|_{N\rightarrow \infty}=\int^{\pi}_{0}[(\frac{1}{\xi_k}(1-\frac{\epsilon_k^2}{\xi_k^2})-\chi_k)+\chi_k]dk,
    \label{eq-Psi_lambda1}
\end{eqnarray}
where $\chi_k = 1/\sqrt{(\delta \lambda)^{2}+(\delta \lambda+\gamma^{2})k^{2}}$. The second part in the above integrand can be computed as,
\begin{equation}
    \int^{\pi}_{0} \chi_k dk = -{1 \over |\gamma|} \ln|\lambda -1| + {\ln(2\pi|\gamma|) \over |\gamma|} + \mathcal{O}(\lambda -1).
\end{equation}
The first integrand in general can not be computed exactly, yet at the critical point ($\lambda = 1$), it can be computed exactly (the expression 
is too complex to be presented here). Gathering all these results together gives 
$\frac{d\Psi_{g}}{d\lambda}|_{N\rightarrow \infty} = \alpha_2 \ln|\lambda -1| + \beta_2 + \cdots$, where
\begin{equation}
    \alpha_2 = -{1 \over |\gamma|}, \quad \beta_2 =  {\ln(8\gamma^2) \over |\gamma|}  - {1 \over |\gamma|},
    \label{eq-a2}
\end{equation}
and the next leading term is $(\lambda - 1)\ln|\lambda - 1| $. These next leading terms attribute to the errors in fitting the constants $\alpha_1$ and
$\alpha_2$ in numerical simulations; and they may become important in the second-order derivative of the GP, see below.

These findings, to the leading orders, are consistent with the numerical results in \cite{zhu2006scaling}. We find that
in these two scaling laws, $\alpha_1 \equiv -\alpha_2$ exactly. Notice that $|\gamma|$ is nothing but just the slope for the closing of energy gap at the critical point, thus 
$\alpha_1$ and $\alpha_2$ is only determined by the inverse of the slope near the critical point, which is the physical meaning of these two constants. The other two 
$\beta$-constants, which are unique functions of $\alpha$, may have the same or opposite sign depending strongly on the value of $\gamma$. 
Moreover, we also have two intriguing limits for these constants. When 
$\gamma \rightarrow \infty$, all these four constants will approach zero in the manner of ${1 \over |\gamma|}$, while on the opposite limit $\gamma \rightarrow 0$, these four 
constants will approach infinity. In both limits, $\alpha_1/\beta_1 \sim 1/\ln(\gamma)\sim 0$ and $\alpha_2/\beta_2 \sim 1/\ln(\gamma) \sim 0$.

This method can also be applied to study the scaling of FS defined as $|\langle g(\lambda)|g(\lambda + d\lambda)\rangle| = 1 - N\Xi_F d\lambda^2/2$ across the critical 
point\cite{gu2010fidelity, gu2004entanglement, gu2010fidelity}. For Eq. \ref{eq-bdg}, we have
\begin{equation}
  \Xi_F  = \frac{1}{4N}\sum_{k>0}(\frac{d\theta_{k}}{d\lambda})^{2} = \frac{1}{4N}\sum_{k>0}\frac{1}{\xi^2_k}(1-\frac{\epsilon^2_k}{\xi^2_k}).
    \label{eq-FS}
\end{equation}
This expression is quite similar to Eq. \ref{eq-Psi_g} except the $(\gamma k)^{-2}$ divergence at the critical point, for which reason the singular function should be
chosen as $\chi_k = {1 \over \gamma^2k^2}$. Similarly we first consider the scaling law as a function of system length $N$, in which the summation of 
$k$ gives 
\begin{equation}
    \frac{1}{4N}\sum_{k>0}\frac{1}{\gamma^{2}k^{2}} = {N \over 96\gamma^2} - {1 \over 8 \pi^2 \gamma^2} + {1 \over 8\pi^2 \gamma^2 N} +\cdots,
\end{equation}
thus $\alpha_1' = 1/(96\gamma^2)$; and the remained part gives
\begin{equation*}
        \frac{1}{8\pi}\int^{\pi}_{0}[\frac{1}{\xi^2_k}(1-\frac{\epsilon^2_k}{\xi^2_k})-\chi_k]dk \approx {1 \over 8\pi^2 \gamma^2} + {\gamma^2 -3 \over 64|\gamma|^3}.
\end{equation*}
Thus we have $\Xi_F|_{\lambda = \lambda_c} = \alpha_1' N + \beta_1'$, where $\beta_1' = {\gamma^2 -3 \over 64|\gamma|^3}$.

\begin{figure}
    \centering
    \includegraphics[width=1.4in]{betaprime.eps}
    \includegraphics[width=1.8in]{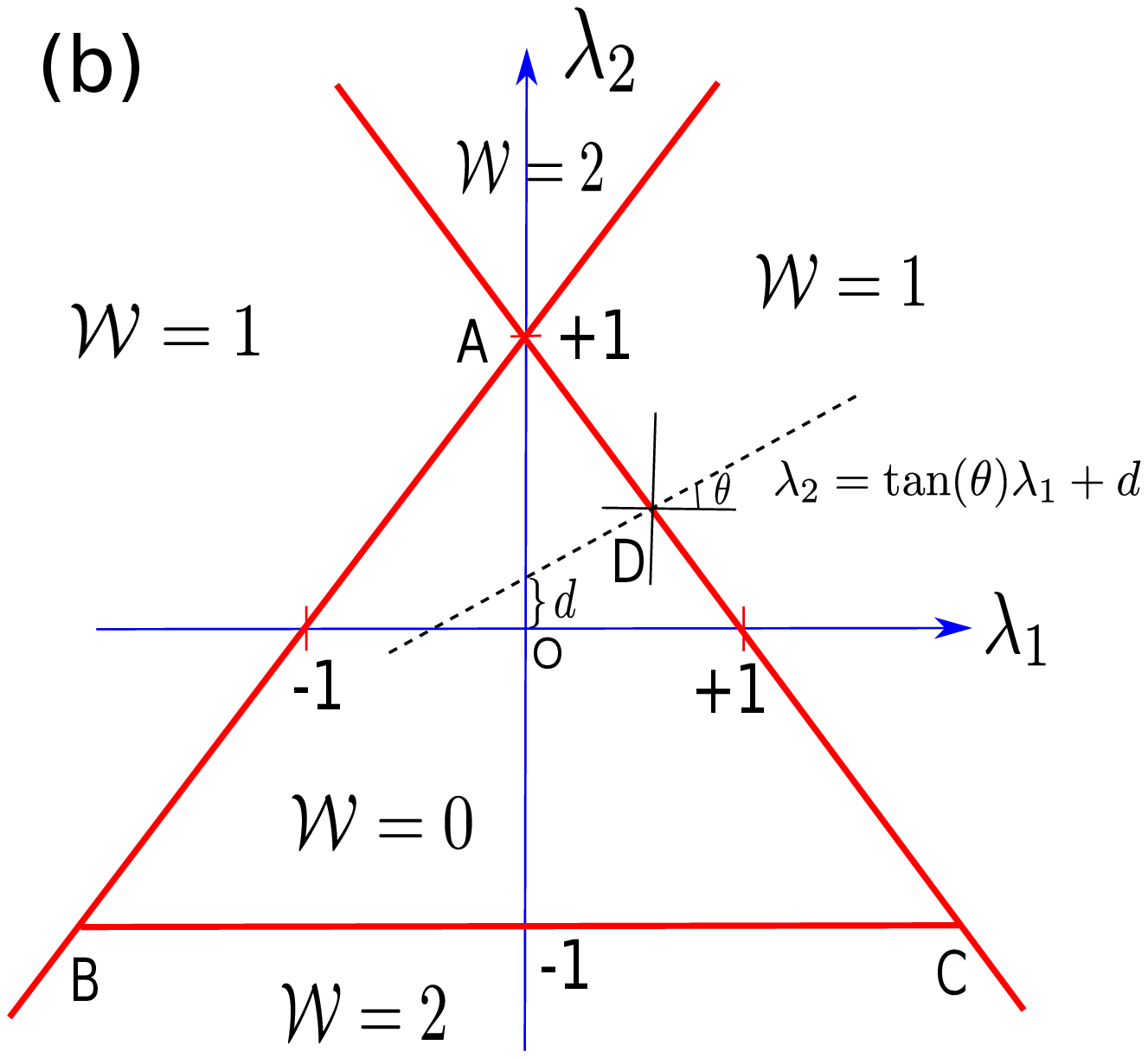}
    \caption{(Color online). (a) Jump of $\beta_2' = d\Psi_g/d\lambda - \alpha_2'/ |\lambda - 1|$ in the anisotropic XY model for 
    $\gamma = 1$, where $\alpha_2'$ is assumed to be a $\lambda$ independent constant. (b) Phase diagram for the extended Ising model, in which 
    different phases are distinguished by winding number $\mathcal{W}$. The condition for the phase boundaries are presented in Eq. \ref{eq-I} to \ref{eq-III}. }
    \label{fig-fig1}
\end{figure}

In the thermodynamic limit, FS as a function of deviation $ \delta \lambda = \lambda - 1$ is computed similarly with the 
singular function $\chi_k = 1/(\delta \lambda^{2}+(\delta \lambda+\gamma^{2})k^{2})$, and we have
\begin{equation}
    \Xi_F=\frac{1}{8\pi}(\int^{\pi}_{0}[\frac{1}{\xi^2_k}(1-\frac{\epsilon^2_k}{\xi^2_k})-\chi_k]dk+\int^{\pi}_{0} \chi_k dk).
    \label{eq-chik}
\end{equation}
Again the second integrand can be computed exactly,
\begin{eqnarray}
    && \frac{1}{8\pi} \int^{\pi}_{0} \xi_k dk =\frac{\text{tan}^{-1}(\pi \sqrt{\lambda -1 + \gamma^2}/ (\lambda -1))}{8\pi(\lambda -1)\sqrt{\lambda -1 + \gamma^2}} \notag \\
 && = {\alpha_2' \over |\lambda -1|} - {1 \over 8\pi^2 \gamma^2} - {\text{sign}(\lambda -1) \over 32|\gamma|^3} + \cdots, 
\end{eqnarray}
where $\alpha_2' = 1/(16|\gamma|)$. The first part can also be computed exactly at the critical point, and finally we find 
\begin{equation}
    \beta_2' = {\gamma^2 -3 \over 64|\gamma|^3} - {1 \over 32|\gamma|^3}\text{sign}(\lambda -1),
    \label{eq-beta2}
\end{equation}
thus we find a different scaling law $\Xi_F|_{N\rightarrow} = \alpha_2'/|\lambda -1| + \beta_2'$, where $\beta_2'$ has a discontinuous jump across the phase boundary owing to 
the appearance of absolute symbol in the denominator\cite{JumpedFunction}; see numerical simulation in Fig. \ref{fig-fig1}a. The jump can be extremely large when $|\gamma|$ is 
small. This jump has been ignored in previous numerical simulations due to its minor role in the divergent behavior of  FS when $\gamma$ is not small enough, thus 
it has been mistakenly declared to be a universal scaling law across the phase boundary. Notice that the jumping of $\beta_2'$ can be absorbed to the singular function by assuming $\alpha_2'$ 
$(= \alpha_2'(\lambda_c) + d\alpha_2'/d\lambda|_{\lambda_c} (\lambda - \lambda_c))$ to be 
$\lambda$-dependent\cite{JumpedFunction}, but this will not change our conclusion. Similar to the discussion before, we find that these two constants approach zero when $\gamma \rightarrow \infty$, 
and infinity when $\gamma \rightarrow 0$. From this result we can also see that the logarithmic divergence in  GP is purely from the linear divergence of the 
expression at the critical point, and some different scaling laws can be found, for instance in the Dicke model\cite{chen2006critical,plastina2006scaling} and Lipkin-Meshkov-Glick 
model\cite{dusuel2004finite,orus2008equivalence,leyvraz2005large} where the gaps are closed in a different ways. Thus these coefficients do not directly carry information of the 
global topology of wave functions.

Coefficients of the divergent terms may be written as,
\begin{equation}
    \alpha_2 = -\alpha_1, \quad \alpha_1' = {1 \over 96} \alpha_1^2, \quad \alpha_2' = {1 \over 16} \vert\alpha_1\vert,
    \label{eq-general}
\end{equation}
which is always correct for a system with gap closed and reopened in a linear way at the special points. The latter two equations also indicate the general relation $\alpha_1' = 
{8 \over 3} (\alpha_2')^2$. Thus these quantities although defined in totally different ways in actually describe the same physics. Besides, from the standard scaling ansatz
\cite{gu2010fidelity, venuti2007quantum,polkovnikov2010universal},
\begin{equation}
    \Xi_F|_{\lambda = \lambda_c} \sim N^{2/\nu -D}, \quad
    \Xi_F|_{N\rightarrow \infty} \sim |\lambda -\lambda_c|^{D \nu -2}.
    \label{eq-scaling}
\end{equation}
For one dimensional system, $D = 1$, our analytical results show the critical exponent for the coherent length $\nu \equiv 1$ exactly. The same conclusion can 
be obtained based on scaling of GP, where $\nu = |\alpha_1/\alpha_2| = 1$. 

When both the scaling of size and parameter are taken into account, we can computer the scaling of $F_1 = {d\psi_g \over d\lambda}|_{\lambda} - {d\psi_g \over d\lambda}|_{\lambda_c}$
and $F_2 = \Xi_F|_{\lambda} - \Xi_F|_{\lambda_c}$ as a function of $N^{\eta} (\lambda - \lambda_c)$. These two scaling functions are only determined by the divergent term 
since $N^{\eta}(\lambda -\lambda_c) \ll |\lambda_c|$. For $F_1$ we find
\begin{equation}
    F_1 = {\pi \over 2} \sum_k {1 \over \sqrt{dx^2 + N^2\gamma^2 k^2}} - {1 \over N|\gamma k|} \simeq \sum_{n=1}^{\infty} {dx^2 \over \mathcal{A}_n},
\end{equation}
where $\mathcal{A}_n = 16n^3 \pi^2 |\gamma|^3 + 3n|\gamma|dx^2$ and $dx = N(\lambda - \lambda_c)$. Notice that the summation of $n$ is extended to infinite due to 
the fast convergence of the above result. After a bit computation we find $F_1 = {\psi_2(1) |N(\lambda - \lambda_c)|^2 \over 32\pi^2|\gamma|^3}$, where 
$\psi_2(1) = -2.40411$ is the polygamma function, thus $\eta = 1$. This result is consistent with the numerical finding in \cite{zhu2006scaling}. The same method can be 
applied to $F_2$ which yields $F_2 = -{|(\lambda - \lambda_c)N^{3/2}|^2 \pi \over 1440|\gamma|^4}$, thus $\eta = 3/2$. 

\textit{Extended Ising model}. We next show that these scaling laws depend strongly on at which point the gap is closed and reopened
and  they may break down when the gap is not closed at the special points, which can be captured by the following extended 
Ising model\cite{niu2012majorana, zhang2015topological, cheng2010fidelity, leyvraz2005large, pachos2004three},
\begin{eqnarray}
  H^{\prime}=-\sum_{j=-M}^{M}(\lambda_{1}\sigma_{j}^{x}\sigma_{j+1}^{x}+\lambda_{2}\sigma_{j-1}^{x}\sigma_{j}^{z}\sigma_{j+1}^{x}+\sigma_{j}^{z}).
\end{eqnarray}
This model can still be exactly solved using the same method\cite{sachdev2007quantum, lieb1961two, pfeuty1970one}. In the fermion picture (Eq. \ref{eq-bdg}) the three-site interaction 
is equivalent to the next-nearest-neighbor hopping and pairing determined by $\lambda_2$, thus we have
\begin{equation}
        i\Delta_k e^{2i\phi} = \sum_{n=1}^2 \lambda_n \sin(nk), \epsilon_k = 1 - \sum_{n=1}^2 \lambda_n \cos(nk).
            \label{eq-EIsing}
\end{equation}
The closing of energy gap determined by $\Delta_k = 0$ and $\epsilon_k = 0$ simultaneously yields
\begin{eqnarray}
    &&  \text{Line AC}: k_0 = 0,   \quad    1 - \lambda_1 - \lambda_2 = 0,  \label{eq-I} \\
    &&  \text{Line AB}: k_0 = \pi, \quad    1 + \lambda_1 - \lambda_2 = 0,  \label{eq-II} \\
    &&  \text{Line BC}: k_0 = \text{cos}^{-1}(\frac{\lambda_1}{2}), \quad  \lambda_2 = -1, |\lambda_1| < 2.
    \label{eq-III}
\end{eqnarray}
The corresponding phase diagram is presented in Fig. \ref{fig-fig1}b. Notice that the BdG equation possesses chiral symmetry $\mathcal{S} = \sigma_x$ at $\phi = 0$, 
where $\sigma_x$ is the Pauli matrix and $K$ is the complex conjugate operator, since $\mathcal{S} H_\text{BdG} \mathcal{S}^\dagger = - H_\text{BdG}$.
This equation belongs to topological BDI class in one spatial dimension\cite{schnyder2008classification, ryu2010topological},  which is characterized by the well-defined 
winding number $\mathcal{W} = {1 \over 2\pi i} \oint dk q^{-1} dq$, where $q = \epsilon_k + \Delta_k$. This model has been studied by Niu et al\cite{niu2012majorana} to show the 
possibility of hosting multiply Majorana fermions in an open chain when $\mathcal{W} =2 $. 

Due to the presence of two parameters in determing the phase boundaries, the divergence of GP and FS depend strongly on how and along which direction the critical boundary is crossed. 
Consider a line across the critical boundary $AC$ along $\theta$ direction (see point $D$ in Fig. \ref{fig-fig1}b and the dashed line is assumed to be $\lambda_2 = \tan(\theta) \lambda_1 + d$)
Then we find the coordinate of $D = ({1-d \over 1+ \tan(\theta)}, {d+\tan(\theta) \over 1+ \tan(\theta)})$. With the previous method we have ($\alpha_2 > 0$ and $\alpha_1 < 0$),
\begin{equation}
    \alpha_2 = - \alpha_1 = + {|1+\tan(\theta)| \over |1 +d + 2\tan(\theta)|},
    \label{eq-k0}
\end{equation}
from which we see that $\alpha_2= -\alpha_1 = \infty$ when $\tan(\theta) =-{d+1\over 2}$; and $\alpha_2 = -\alpha_1 =0$ when along the phase boundary ($\theta = -\pi/4$ or $3\pi/4$) 
since no phases are crossed. When $\theta = \pi/2$, we have $\alpha_2 = -\alpha_1 \equiv {1 \over 2}$, which is independents of the other parameters. The other two coefficients can also 
be defined straightforwardly using Eq. \ref{eq-general}.
The constants $\beta_1$ and $\beta_2$ in this extended model can no longer be computed analytically, however, they can still be computed exactly with the technique in
Eq. \ref{eq-Psi_lambda1} and \ref{eq-chik} using numerical methods.

\begin{figure}
    \centering
    \includegraphics[width=0.5\textwidth]{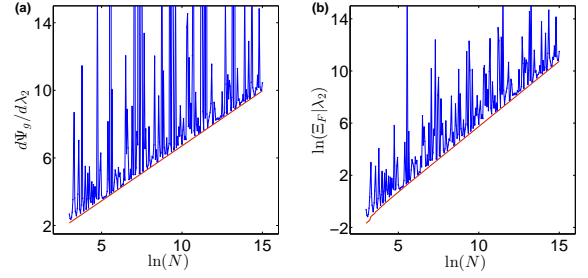}
    \caption{(Color online). Breakdown of scaling laws for (a) ${d\Psi_g\over d\lambda_2}$ and (b) $\Xi_F|_{\lambda_2}$ as a function of length across
    the boundary $BC$. We set $\lambda_1=\frac{\sqrt{5}}{2}+\frac{2}{5}$ to avoid the singular point $k_0$. The solid line in (a) is $\alpha_1 \ln(N) + \beta_1$ 
    and in (b) is $\alpha_1' N + \beta_2'$, where $\alpha_1 = |\sin(k_0)|$ and $\alpha_1' = {1 \over 192}$ are determined using singular function expansion method. 
    The two $\beta$-constants are fitted using the lowest bound of the data.}
    \label{fig-fig2}
\end{figure}

Along the boundary $BC$, we find ${d\psi_g \over d\lambda_1} \propto (1+\lambda_2)$ and $\Xi_F|_{\lambda_1} \propto (1+\lambda_2)$, thus both ${d\psi_g \over d\lambda_1} = 0$ and 
$\Xi_F|_{\lambda_1} = 0$ exactly as a function of length $N$ and deviation $\delta \lambda$ for the same reason. We next point out that the scaling laws as a function of $N$ 
across the phase boundary $BC$ along $\lambda_2$ direction is broken-down since the gap is not closed and reopened at the special points. This is different from 
the previous case where $k_0 = 0$ or $k_0 = \pi$ will not be sampled during the summation of $k$. A typical result for the GP and FS is presented in Fig. \ref{fig-fig2}, in which 
we find that at some ''magic point'' when $k = {2\pi n/N} \rightarrow k_0$, a ''pulse'' in these two quantities 
can be found. The analogous features can also be found in other extended models\cite{zhang2015topological,leyvraz2005large, pachos2004three, cheng2010fidelity}. The breakdown of 
this scaling also indicates the failure of Eq. \ref{eq-scaling} and scaling of ${d\psi_g \over d\lambda}|_{\lambda_2} - {d\psi_g \over d\lambda}|_{\lambda_{2c}}$ and 
$\Xi_F|_{\lambda_2} - \Xi_F|_{\lambda_{2c}}$ as a function of $N^{\eta}(\lambda_c - \lambda_{2c})$.

However, the similar scaling laws can still be found as a function of deviation $\delta \lambda = \lambda_2 +1$ (line $BC$). In the vicinity of 
$k_0$ the energy gap can be approximated as,
\begin{equation}
    \xi_k^2 \approx a + b (k-k_0) + c(k-k_0)^2,
\end{equation}
where $c = 2-2\cos(2k_0)$, $b = 2 \delta \lambda \sin(2k_0)$ and $a=\delta\lambda^2$, with $\cos(k_0) = \lambda_1/2$ and $b^2 -4ac \le 0$ for $\forall$ $k$. 
This series expansion is different from the 
previous ones due to the appearance of linear term $b$. Notice that when $k_0 = 0$ or $\pi$, the contribution of the numerator in the integrand is always equals to one; 
however in this case, the numerator then becomes important, and the singular function should be chosen as $\chi_k = {\sin^2(k_0) \over \xi_k}$ in GP and $\chi_k ={\sin^2(k_0) \over \xi^{2}_k}$ 
in FS. Thus the final coefficients $\alpha_i$ are no longer purely determined by the slope $c$. With these singular function expansions we find 
\begin{eqnarray}
    &&    {d\Psi_g \over d\lambda_2} = \alpha_2 \ln |\lambda_2+1| + \beta_2,  
    \Xi_{F}\vert_{\lambda_2} = {\alpha_2' \over |\lambda_2+1|} + \beta_2',  \\
    && \alpha_2 = -{2 \over \sqrt{4 -\lambda_1^2}}, \alpha_2' = {\pi/2  - \text{tan}^{-1}(\lambda_1/\sqrt{4 -\lambda_1^2}) \over 8\pi}.
\end{eqnarray}
The intimate relations in Eq. \ref{eq-general} due to the contribution of the numerator at non-special $k_0$ is no longer true (it still holds only when $\lambda_1=0$). 
The above result is correct only when $\lambda_1$ is not very close to $\pm 2$ (points $B$, $C$), in which case the constants $\beta_2$ and $\beta_2'$ may become 
singular. 

{\it Discussion and Conclusion.} Here a general method to obtain the exact scaling laws for GP and FS across the quantum phase transitions is presented. These scaling laws are independent of the 
choice of singular functions since for different singular functions the divergent behavior near the critical points which determine the scaling laws are exactly the same. 
Moreover this method can be applied not only to the first-order derivative of GP but also their higher-order derivatives across the critical point. For instance, for Eq. \ref{eq-XY},
\begin{equation}
    {d^2\Psi_g \over d\lambda^2}|_{\lambda =1} = -{3 \over 2|\gamma|^3} \ln N + \beta_3,
\end{equation}
with $\beta_3 = \frac{3 (\ln\frac{\pi}{\vert\gamma\vert}-(\Gamma +3 \ln2 - 4))} {2|\gamma|^3} - \frac{1}{2\vert\gamma\vert}+\cdots$, 
which has the same form as ${d\Psi_g \over d\lambda}|_{\lambda =1}$. However for the deviation $\delta \lambda = \lambda -1$, it takes another intriguing form after
singular function expansion,
\begin{eqnarray}
    {d^2\Psi_g \over d\lambda^2}|_{N\rightarrow \infty} 
    = -{1\over |\gamma| (\lambda -1)} + {3 \ln(|\lambda-1|) \over 2|\gamma|^3} + \beta_3^\prime,
    \label{eq-beta3}
\end{eqnarray}
where $\beta_3^\prime=(3\ln\frac{\pi^2}{2}+4)/(2\vert\gamma\vert^3)-1/(2\vert\gamma\vert)$. These two singular functions arise from the derivative 
of the leading term, $\ln |\lambda -1|$, and the next leading term, $\frac{3}{2\vert\gamma\vert^2}(\lambda -1)\ln(|\lambda -1|)$. 
The jumping of constant $\beta_3'$ is absent\cite{JumpedFunction}.

These results also reveal a close relation between the constants $\beta$ and $\alpha$. These interesting features, which have been barely discussed in previous literatures, will be presented elsewhere. 
This method is powerful and can also be adapted to study the singular behaviors in entanglement\cite{osterloh2002scaling, gu2003entanglement, osborne2002entanglement,gu2004entanglement,
lambert2004entanglement,vidal2003entanglement, wu2004quantum} quantum discord and correlation\cite{dillenschneider2008quantum, sarandy2009classical, chen2010quantum} and geometric Euler 
number\cite{venuti2007quantum, yang2015geometric,ma2013euler}, which will be subject to future investigation. To conclude, these exact results can greatly enrich our understanding of GP and FS in the characterization of quantum phase transitions.

{\it Acknowledgement.} M.G. is supported by the National Youth Thousand Talents Program (No. KJ2030000001), the USTC start-up funding (No. KY2030000053) and the CUHK RGC Grant (No. 401113). Z. Z. and G. G. are supported by National Key Research and Development Program (No. 2016YFA0301700), National Natural Science Fundation of China(No. 11574294), and the ''Strategic Priority Research Program (B)'' of the Chinese Academy of Sciences (No. XDB01030200).

\begin{acknowledgments}

\end{acknowledgments}



\bibliography{berryphase_spin_model}  

\begin{thebibliography}{77}
\bibitem [{Jum()}]{JumpedFunction}%
  \BibitemOpen
  \href@noop {} {}\bibinfo {note} {For a general singular function,
  $\mathcal{G}(\lambda) = {f(\lambda) \over |\lambda - \lambda_c|}$, where
  $f(\lambda)$ is smooth function of $\lambda$, its scaling law near the
  critical point can be written as $\mathcal{G}(\lambda) = {f(\lambda_c) \over
  |\lambda - \lambda_c|} + f'(\lambda_c)\text{sign}(\lambda - \lambda_c) +
  \text{finite}$, which explains the jumping of constant $\beta_2'$ in Eq.
  \ref{eq-beta2}. We do not have the jumped constant in Eq. \ref{eq-beta3} due
  to the absence of absolute symbol in the denominator.}\BibitemShut {Stop}%
\end{thebibliography}

\end{document}